\documentstyle[aps]{revtex}

\begin{document}

\title{High Resolution $^{13}C$ NMR study of oxygen intercalation in $C_{60}$}
\author{P. Bernier$^1$, I. Luk'yanchuk$^{1,2}$, Z. Belahmer$^1$, M. Ribet$^1$,
L. Firlej$^1$}
\address{$^1$
Groupe de Dynamique des Phases Condens\'ees, Universit\'e des Sciences et
Techniques du Languedoc, \\ 34060 Montpellier, France}
 \address{$^2$L.D.Landau
Institute for Theoretical Physics, Moscow, Russia }

\date{\today}
\maketitle

\widetext

\begin{abstract}
Solid state high resolution $^{13}C$ NMR has been used to investigate the
physical properties of pristine $C_{60}$ after intercalation with molar
oxygen. By studying the dipolar and hyperfine interactions between Curie
type paramagnetic oxygen molecules and $^{13}C$ nuclei we have shown that
neither chemical bonding nor charge transfer results from the intercalation.
The $O_2$ molecules diffuse inside the solid $C_{60}$ and occupy the
octahedral sites of the fcc crystal lattice. The presence of oxygen does not
affect the fast thermal reorientation of the nearest $C_{60}$ molecules.
Using  Magic Angle Spinning we were able to separate the dipolar and
hyperfine contributions to $^{13}C$ NMR spectra, corresponding to buckyballs
 adjacent to various numbers of oxygen molecules.
\end{abstract}

\narrowtext
\setlength{\parindent}{5pt} \leftskip -10pt \rightskip 10pt

\section{INTRODUCTION}

$C_{60}$ under its molecular or solid form, is an ideal system for $^{13}C$
NMR investigation. All carbons on the molecule are chemically equivalent,
yielding an unique resonance at 143.6 ppm from the TMS reference,
characteristic of an aromatic system \cite{1}. Also, due to the Van der
Waals nature of the solid, intermolecular interactions are very weak and the
resonance in the solid is similar to that in solution \cite{2}. Such a
situation is particularly interesting as any change on the ball itself
(substitution, addition) or any chemical intercalation in the fcc structure
of the solid will be easily detected by $^{13}C$ NMR.

A simple situation is observed when the intercalant does not induce charge
transfer with the host molecules. In this case the measured shift for the $%
^{13}C$ resonance is a chemical shift eventually corrected by contributions
coming from the environmental disturbance. A typical case is $C_{60}$
intercalated with oxygen. As first shown by Assink et al \cite{8}, by
submitting $C_{60}$ pr to a high pressure of oxygen, this species easily
diffuses in the octahedral sites of the fcc structure. Due to the
paramagnetic nature of the molecule, the contact interaction between $^{13}C$
and oxygen orbitals produces a small shift whose amplitude is proportional
to the magnetization (Curie law). This effect is additive and the shift
corresponding to $C_{60}$ molecules, surrounded by $q$ $(6\geq q\geq 2)$
oxygens is $q$ times larger than the shift caused by one oxygen. Assink et
al have also shown that the $^{13}C$ relaxation time for $C_{60}$ molecules
surrounded by one or more oxygens is considerably decreased if compared with
pure $C_{60}$ \cite{8}. More recently, Belahmer et al have shown \cite{9}
that all these effects occur even if only ambient oxygen pressure conditions
are imposed to the system.

In this paper we present a new study of the effect of oxygen. By comparing
High-Resolution and static NMR spectra we were able to pick out the
dipole-dipole interaction between the $^{13}C$ nuclear and $O_2$ magnetic
moments. This allowed us to calculate the $C_{60}$-$O_2$ intermolecular
distance very accurately and to conclude that oxygen molecules carry the
electronic spin $S=1$ and occupy only the octahedral sites of the fcc
lattice. No influence of the intercalated oxygen on the chemical and
dynamical properties of $C_{60}$ has been observed. This conclusion was
confirmed by the investigation of the dipolar contribution in the case of
two surrounding oxygens, extracted from the sidebands envelope of the Magic
Angle Spinning spectra.

\section{EXPERIMENTAL}

The $C_{60}$ powder has been obtained using a standard procedure yielding a
99.9\% pure sample. The whole batch was kept in air and in the dark for a
period of at least four months prior to experiment. In some cases a 9\% $%
^{13}C$ enriched sample was used in order to improve thenal to noise
ratio.

$^{13}C$ NMR spectra of powdered highly crystalline samples were recorded on
a Bruker CXP200 or ASX200 spectrometer working at 50.3 MHz. Magic angle
spinning (MAS with a rotational frequency from 0.1 to 6kHz) was necessary to
obtain high resolution and sensitivity. The chemical shifts are reported to
the classical reference TMS.

Variable temperature (in the range 220 to 350K) with high resolution
conditions was obtained using a cold or hot bearing gas flow

\section{RESULTS AND DISCUSSION}

Figure 1 presents the spectrum obtained at room temperature and ambient air
pressure with MAS (at 4 kHz) for the 9\% $^{13}C$ enriched sample. In
addition to the expected resonance at 143.6 ppm due to $C_{60}$ molecules in
a perfect fcc environment, we observe five weak resonances whose positions
are multiple of + 0.7 ppm from the main resonance. By reference to the work
performed by Assink et al. \cite{8} it is clear that these resonances
correspond to the various cases of oxygen occupancy in the six octahedral
sites around each $C_{60}$ molecule (Figure 2). Note that in our case we do
not observe the sixth line which is probably too weak as we work at ambient
air pressure. The first resonance close to the main line then corresponds to
the case when one oxygen is adjacent to a $C_{60}$ molecule, the second peak
corresponds to the two oxygen environment etc.

It must be noticed that the intensities of these lines rapidly decrease and
scale as $I_1:I_2:I_3:I_4:I_5=1:0.04:0.007:0.005:0.001$. The ratio $%
I_0:I_1=0.025$ obtained at the condition of fully relaxed pure $C_{60}$
spectrum with intensity $I_0$ with repetition time of $5min$, gives the
concentration of oxygen as $n=2.5\%$ per $C_{60\text{ }}$ molecule. Compare
now the factor of the ratio of the first two peaks, $I_1:I_2=1:0.04$, with
that which would be expected in the case when 2.5\% of $O_2$ molecules
were uniformly distributed over the sample. The later is provided by the
probability of having two $O_2$ molecules located near the same buckyball
sphere and is given by: $I_1:I_2=18n^2=1:0.01$ where factor $18$ is the
number of possible octahedral positions which can be joined with the given
oxygen octahedral site {\it via} neighboring $C_{60}$ molecules. Since this
ratio is smaller than that  measured experimentally we conclude that the
equilibrium condition of the uniform oxygen distribution is not achieved and
oxygen concentration diminishes from the surface of powder pieces to their
center. Note however that in our case the oxygen distribution is much close
to the uniform one than it was observed in the experiments of Assink\cite{8}
where the characteristic values of $I_1:I_2$ were $1:0.4\div 1:0.5$ with
approximately the same oxygen concentration and the exposition time was
measured in hours.

{}From the $^{13}C$ NMR spectrum of oxygen doped fullerenes we argue that no
charge transfer occurs from $O_2$ molecules to $C_{60}$. All the spectrum
properties are provided by the paramagnetic ground state of molecular
oxygen, where two valence electrons form the triplet state with $S=1$.  The
shift is attributed\cite{8} to the isotropic hyperfine coupling, which
presumably has the Fermi-contact type, of $^{13}C$ nuclei with the Curie
type magnetic moments $\mu _S$ of the neighbor oxygen molecules: $\mu
_S=(g\mu _B)^2S(S+1)B_0/3kT=4\mu _B{}^2S(S+1)B_0/3kT$ (with $g=2$).

The hyperfine interaction, being strongly dependent on the distance between $%
\mu _S$ and $^{13}C$ nuclei, is effectively averaged by the fast thermal
reorientation of those and therefore can be described by the hyperfine
coupling constant $A_{eff}$ taken as an average of the $\mu _S-^{13}C$
hyperfine interaction over the neighbor to the oxygen buckyball re. The
isotropic shift, $\sigma _{iso}$, is then inversely proportional to the
temperature and to the number of oxygens $q$ (with a maximum of 6) neighbors
of a given $C_{60}$ molecule:

\begin{equation}
\sigma _{iso}=q\left( \frac{A_{eff}}\hbar \right) \frac{g\mu _BS(S+1)}{%
3kT\gamma _I}  \label{1}
\end{equation}

where $\gamma _I$ is a nuclear magnetogyric ratio. The inverse temperature
behavior of $\sigma _{iso}$ observed experimentally\cite{8}\cite{Zineb}%
confirms the hyperfine rather than charge transfer origin of $\sigma _{iso}$.

By studying the anisotropic spectrum obtained with no MAS we show that the
intercalated oxygen molecules do have the spin $S=1$ and therefore no
electrons came from $O_2$ to $C_{60}$. Figure 3(a) presents the static
spectrum obtained with such a condition on the same sample as above. A
superposition of two components is observed:

i) a symmetric line centered at 143.6 ppm, with a width of roughly 3 ppm,
which correspond to $C_{60}$ molecules not ajucent to any oxygen molecule;

ii) an anisotropic spectrum which expands from 125 ppm to 175 ppm and
corresponds to $C_{60}$ molecules ajucent to only one oxygen. Taking into
account the intensities of the various isotropic resonances (Fig.1), we
expect that the anisotropic spectrum corresponding to two or more oxygens
should be extremely weak. Note also that at very low spinning frequency
(MAS) of the sample (Fig.3(b)), all the observed spinning side bands
correspond to the isotropic resonance of the ''one oxygen'' case, and their
envelope reproduces the static spectrum of Figure 3(a) \cite{10}.

The origin of the anisotropic spectrum is the IS dipolar interaction between
the oxygen and $^{13}C$  magnetic moments. Two limiting situations
have to be considered: the $O_2$-$^{13}C$ vector $r_i$, can be either
parallel or perpendicular to the external magnetic field. Ie first case
the shift $\sigma _{\Vert }$ is paramagnetic, while in the second one the
shift $\sigma _{\bot }$ is diamagnetic and twice as smaller in magnitude as $%
\sigma _{\Vert }$. The powder spectrum is provided by the all intermediate
positions of the vector $r_i$.

On a quantitative basis, the IS dipolar interaction results in the traceless
uniaxial NMR shift tensor $\sigma _{ij}$:

\begin{equation}
\sigma _{ij}=\mu _S\left( 3\frac{r_ir_j}{r^5}-\frac{\delta _{ij}}{r^3}%
\right) \frac 1{B_0}  \label{2}
\end{equation}

The fast thermal reorientation of $C_{60}$ molecule leads to the effective
averaging of (2) over all the possible locations of $^{13}C$, i.e. over the
surface of the neighboring $C_{60}$ molecule. Calculating this average
finally we obtain:

\begin{equation}
\overline{\sigma }_{ij}({\bf l})=\frac{4\mu _B}{3kT}S\left( S+1\right)
\left( 3\frac{l_il_j}{l^5}-\frac{\delta _{ij}}{l^3}\right)  \label{3}
\end{equation}

Where $l_i$ is the vector joining the centers of the $O_2$ and $C_{60}$
molecules. The static NMR spectrum is determined by the powder pattern of
the tensor and has the shape shown on Figure 3. The principal values of $%
\overline{\sigma }_{ij}$, which corresponds to $\overline{\sigma }_{\bot }$
and $\overline{\sigma }_{\Vert }$ give the width of the spectrum:$\Delta
\overline{\sigma }=\overline{\sigma }_{\Vert }-\overline{\sigma }_{\bot }$ .
{}From (3) we have:
\begin{eqnarray}
\overline{\sigma }_{\bot }=-S\left( S+1\right) \frac 4{3kT}\frac{\mu _B^2}{%
l^3}\qquad \overline{\sigma }_{\Vert }=2S\left( S+1\right) \frac 4{3kT}\frac{%
\mu _B^2}{l^3}  \nonumber \\
\Delta \overline{\sigma }=\frac 4{kT}S\left( S+1\right) \frac{\mu _B^2}{l^3}%
\qquad  \label{4}
\end{eqnarray}

Figure 4 shows that the inverse temperature behavior for $\Delta \overline{%
\sigma }$ does take place at least in the range 240-340 K. Note the
cleaobserved drop of $\Delta \overline{\sigma }$ at a temperature close to the
first order transition temperature Tc = 260 K \cite{11}\cite{12}.

{}From the expression above, with $\Delta \overline{\sigma }=48ppm$ at T =
300K, assuming $S=1$,one obtain $l=7\AA $\ (with 4\% accuracy) which exactly
corresponds to the distance from the buckyball sphere center to the nearest
octahedral position of oxygen. We then conclude that:

(i) oxygen molecules stay exact in the middle of the octahedral site of fcc
lattice, no tetrahedral positions are populated by oxygen;

(ii) the oxygen molecules carry the electronic spin $S=1$ and therefore
neither charge transfer nor chemical bounding resulting from oxygen occurs
in the system

(iii) the width of the one-oxygen static NMR spectrum is provided by the
dipolar $^{13}C-O_2$ interaction, rather then the buckyball rotation
hindrance by the intercalated oxygen as it was suggested by \cite{9} . The
intercalated oxygen has no influence on the $C_{60}$ thermal reorientations,
at least in NMR time scale $2 \cdot 10^{-8}s$.

With $l$ being half of the fcc lattice constant, we can imagine that the
small drop of $\Delta \overline{\sigma }$ at 260 K could be due to the
experimentally observed contraction of the lattice at the transition \cite
{11}\cite{12}. Nevertheless this contraction induces a variation of the
lattice parameter smaller than 0.2\%, which cannot account for the measured
16\% drop of $\Delta \overline{\sigma }$ at 260K. We suggest that the large
drop of $\Delta \overline{\sigma }$ could be attributed to the partial
blocking of the movement of the nearest to the oxygen $C_{60}$ molecules
after the transition.

Now we show that the two-oxygen NMR spectrum is also provided by the spin
paramagnetism of oxygen molecules and confirms the given above conclusions
about the oxygen state in $C_{60}$. Technically this cis more delicate to
study. The isotropic part of the spectrum is easy to detect at 1.4 ppm from
the main line (Figure 1), which is twice the value for the one-oxygen case.
This corresponds to the doubling of the effective hyperfine interaction,
according to (1) (with $q=2$). The anisotropic spectrum cannot be obtained
directly as it was in the one-oxygen case. The main reason is the very weak
signal to noise ratio as a result of the small number of $C_{60}$ molecules
surrounded by two oxygens and the increasing width of the anisotropic
spectrum. The only way to get some information on the shape of the spectrum
is to record the High Resolution spectrum with enough spinning side bands to
be able to study their envelope. After this, assuming that the envelope has
a similar form than static spectrum [10] we can compare it to the
theoretical predictions.

The calculation of the shape of the two-oxygen static NMR spectrum uses the
same assumptions as above: the $C_{60}$ balls are supposed to rotate freely
at room temperature and the molecular oxygen occupies the octahedral sites
only. Then we have to consider two cases: two oxygens are located along the
same axis going through the $C_{60}$ center, e.g. A and B on Figure 2
(diametral arrangement) or they are located on the perpendicular axis: A and
C, D, E or F on Figure 2 (rectangular arrangement). Considering a given
octahedral oxygen site, surrounded by six $C_{60}$ molecules, we find that
there exist twelve neighbor octahedral sites which correspond to the
rectangular oxygens arrangement and six corresponding to the diametral case.
Therefore the rectangular location is twice as probable as the diametral one.

Formally, the NMR shift tensor results from the contributions of the dipolar
NMR shift tensors from both oxygens, surrounding $C_{60}$ molecule:

\begin{equation}
\sigma _{ij}^{2ox}\left( {\ b}_1\symbol{94}{\bf l}_2\right) =\overline{%
\sigma }_{ij}({\bf l}_1)+\overline{\sigma }_{ij}({\bf l}_2)  \label{5}
\end{equation}

Here ${\bf l}_1$, ${\bf l}_2$ are the unit vectors directed from the center
of $C_{60}$ to the oxygen positions (either diametral or rectangular
arrangement), $\overline{\sigma }_{ij}({\bf l})$ is the dipolar shift tensor
given by (3). Therefore the relative location of $O_2$ molecules with
respect to $C_{60}$ is important and the intensity of the resulting
lineshape is the sum of two contributions:

\begin{equation}
I^{2ox}=6\cdot I^\pi \left( \sigma \right) +12\cdot I^{\pi /2}\left( \sigma
\right)  \label{6}
\end{equation}

The lineshapes $I^\pi \left( \sigma \right) ,I^{\pi /2} \left( \sigma \right)
$, for the diametral and rectangular oxygen arrangements correspond to the
powder patterns of the tensor (3) with ${\bf l}_1\symbol{94}{\bf l}_2=\pi $
and ${\bf l}_1\symbol{94}{\bf l}_2=\pi /2$ respectively, factor 6 and 12
being the above mentioned population ratios for the diametral and
rectangular oxygen locations. Both oxygen arrangements give powder
lineshapes having uniaxial tensors $\overline{\sigma }_{ij}^{2ox}\left( \pi
\right) $ $\overline{\sigma }_{ij}^{2ox}\left( \pi /2\right) $,with the
following principal values:

{\it for diametral location:}

\begin{eqnarray}
\overline{\sigma }_{\bot }^{2ox}\left( \pi \right) =-\frac{16}{3kT}\frac{\mu
_B^2}{l^3}\qquad \overline{\sigma }_{\Vert }^{2ox}\left( \pi \right) =2\frac{%
16}{3kT}\frac{\mu _B^2}{l^3}\qquad \   \nonumber \\
\Delta \overline{\sigma }^{2ox}\left( \pi \right) =\frac{16}{kT}\frac{\mu
_B^2}{l^3}  \eqnum{7a}
\end{eqnarray}

{\it for rectangular location:}

\begin{eqnarray}
\overline{\sigma }_{\bot }^{2ox}\left( \frac \pi 2\right) =\frac 8{3kT}\frac{%
\mu _B^2}{l^3}\qquad \overline{\sigma }_{\Vert }^{2ox}\left( \frac \pi 2%
\right) =-2\frac 8{3kT}\frac{\ B^2}{l^3}\qquad  \nonumber \\
\Delta \overline{\sigma }^{2ox}\left( \frac \pi 2\right) =\frac 8{kT}\frac{%
\mu _B^2}{l^3}  \eqnum{7b}
\end{eqnarray}

The resulting NMR spectrum extends over more than 90 ppm and presents two
well defined maxima as shown on Figure 5. We have also presented the
experimental amplitudes of the spinning side bands (the accuracy is not
better than 30\% in this case) as obtained after a long accumulation and
using a low spinning frequency (400 Hz). Taking into account the extent of
the spectrum and the poor signal over noise ratio we consider that there is
a reasonable agreement with the predicted spectrum.

\section{CONCLUSION}

Using static and high resolution $^{13}C$ NMR, we have shown that presence
of oxygen in solid $C_{60}$ in equilibrium with ambient air pressure can be
easily detected. The molecular oxygen is located exactly in the middle of
the octahedral sites of the fcc crystal lattice and has the electronic spin $%
S=1$, corresponded to the paramagnetic moment of $O_2$ molecule. Therefore
no charge transfer results from the oxygen intercalation. The case of one
oxygen per $C_{60}$ can be analyzed quite accurately by separate studying of
the dipolar- and contact hyperfine contributions to the NMR spectra. We
conclude that the presence of molecular oxygen does not affect significantly
either the chemical or the dynamical properties of solid $C_{60}$.
Two-oxygen NMR spectra confirm the given conclusions.

{\bf ACKNOWLEDGMENTS}

The Groupe de Dynamique des Phases Condens\'es is Unit\'e de Recherche
Associ\'ee au CNRS n 233. We thank J.E.Fischer for stimulating discussions.
I.L. is greateful to the regional administration of the province Languedoc,
France for financial support (28PAST8016)


 \begin{figure}[tbp]
\caption{$^{13}C$ NMR spectrum of $C_{60}$ powder contaminated with oxygen
at ambient conditions. The intensity of the main line at 143.6 ppm is
artificially decreased by reference to the other lines, as the recycling
time is very short (1.5 s) compared to its relaxation time (at least 100 s).}
\label{Figure 1.}
\end{figure}

\begin{figure}[tbp]
\caption{ Schematic representation of the six octahedral sites (noted A, B,
C, D, E and F) of the fcc lattice around $C_{60}$ molecule.}
\label{Figure 2.}
\end{figure}

\begin{figure}[tbp]
\caption{$^{13}C$ NMR spectra of the sample of Figure 1.
a) Static spectrum
obtained with no MAS showing the superposition of the two lines.
b) Same
conditions, but obtained with a low frequency MAS of 116 Hz.}
\label{Figure 3.}
\end{figure}

\begin{figure}[tbp]
\caption{ Variation of  $\Delta \overline{%
\sigma }$  defined as the distance in ppm between the two sides
of the anisotropic spectrum such as presented in Figure 3. versus the
inverse temperature, showing the linear dependence and the slight drop close
to T = 260 K.}
\label{Figure 4.}
\end{figure}

\begin{figure}[tbp]
\caption{ Amplitudes of the spinning side bands corresponding to the two
oxygens per $C_{60}$ case. We compare the experimental data with the
predictions of theory (full line). The two component spectra (dashed lines)
correspond to the two situations described in the text.}
\label{Figure 5.}
\end{figure}

\end{document}